# Artificial Intelligence-Guided PET Image Reconstruction and Multi-Tracer Imaging: Novel Methods, Challenges, And Opportunities

Movindu Dassanayake[1], Alejandro Lopez[2], Andrew Reader[3], Gary J.R. Cook[3], Clemens Mingels[2,4], Arman Rahmim[5,7], Robert Seifert[2], Ian Alberts*[6], Fereshteh Yousefirizi[7]

1. School of Biomedical Engineering & Imaging Sciences, Kings College London, London, United Kingdom
2. Department of Nuclear Medicine, Inselspital, Bern University Hospital, University of Bern, Bern, Switzerland
3. King's College London and Guy's & St Thomas' PET Centre, School of Biomedical Engineering and Imaging Sciences, King's College London, London, United Kingdom
4. Department of Radiology, University of California Davis, Sacramento, CA, USA
5. Department of Radiology, University of British Columbia, Vancouver, BC, Canada
6. Molecular Imaging and Therapy, BC Cancer, Vancouver, BC, Canada
7. Department of Integrative Oncology, BC Cancer Research Institute, Vancouver, BC, Canada

**Corresponding author:**

Ian Alberts, MA MBBS MD PhD

Molecular Imaging and Therapy

600 West 10th Ave

V5Z 4E6

Vancouver, British Columbia, Canada

ian.alberts@ubc.ca

**Clinical care points**

•       LAFOV PET/CT has the potential to unlock new applications such as ultra-low dose PET/CT imaging, multiplexed imaging, for biomarker development and for faster AI-driven reconstruction, but further work is required before these can be deployed in clinical routine.

•       LAFOV PET/CT has unrivalled sensitivity but has a spatial resolution of an equivalent scanner with a shorter axial field of view. AI approaches are increasingly explored as potential avenues to enhance image resolution.

# Introduction

Positron emission tomography/computed tomography (PET/CT) imaging is an essential tool in oncology, crucial for evaluating tumor biology, staging diseases accurately, and monitoring responses to treatment (1, 2). While metrics like SUVmax have traditionally been used extensively, they face challenges related to consistency and predictive accuracy (3, 4). Recent advancements in artificial intelligence (AI) have significantly enhanced PET image reconstruction and quality.

The emergence of total-body (5) and long-axial field-of-view (LAFOV) (6) PET/CT scanners has significantly expanded these capabilities. Compared to standard axial field-of-view (SAFOV) scanners, LAFOV systems offer substantially extended axial coverage to simultaneously image larger body regions as well as increased sensitivity, up to an order of magnitude higher (7, 8), providing opportunities for enhanced clinical performance and novel research applications (9). These advancements allow for significant reductions in either injected radiotracer activity or shortened scan times(10), and innovative imaging techniques such as multiplexed multi-tracer studies (11). Furthermore, LAFOV PET/CT might allow for ultra-low dose imaging and safe inclusion of healthy cohorts(12), establishing robust physiological baselines critical for disease anomaly detection(13). Furthermore, although dynamic whole-body multi-parametric PET imaging has recently been adopted in SAFOV PET systems employing multi-pass multi-bed WB scan protocols (14, 15), the advent of LAFOV simplified these scan protocols by enabling their application at a single stationary bed position for streamlined, continuous and comprehensive graphical analysis and multi-compartmental tracer kinetic analyses to complement semi-quantitative static SUV-based metrics (16-20).

Building upon advances in AI developed for SAFOV imaging, this review highlights recent developments in AI-driven PET image reconstruction and quantification techniques, with a particular focus on advancements, opportunities, and challenges presented by LAFOV PET/CT imaging.

# Artificial intelligence for image reconstruction: single-tracer and simultaneous multi-tracer PET

The use of AI in PET image reconstruction has been increasingly investigated in recent years (21-24). In SAFOV PET, initially applied techniques involve direct AI methods where the PET system matrix is learnt via a convolutional neural network (CNN) or a similar neural network model at the expense of many training pairs (up to 200,000 2D image pairs (21)). These direct methods provide high quality reconstructions when the data are within the training distribution. More recently, unrolling of iterative methods to incorporate aspects of PET physics into the learning process have shown improvement in generalizability by requiring fewer training samples (in the order of tens of 3D data/images (21)). Another paradigm of AI methods involves purely self-supervised training which does not require training data and could offer the best generalizability but potentially at the cost of image quality (25). With the recent introduction of LAFOV PET some AI methodologies originally introduced for SAFOV PET have been applied and extended. In this section, we review AI methods that have been developed for PET reconstruction related to LAFOV PET. In addition, with the growing need of disease specific imaging, we also cover AI methods for multiplexed imaging in PET where multiple tracers are injected and imaged simultaneously.

*Deep learned PET reconstruction*

FastPET was developed for the Siemens Biograph Vision scanner, employing ordered-subset expectation maximisation (OSEM) reconstructions as labelled targets, and involving training and validating a neural network (NN) with histo-images and attenuation images as inputs (26). This reconstruction method, essentially an image-to-image mapping, significantly accelerates reconstruction compared to conventional OSEM without compromising image quality. This method was later extended to the Biograph Vision Quadra LAFOV scanner (Siemens Healthineers), achieving a reconstruction time reduction from 7 minutes to 20 seconds post-scan, with an average absolute image difference of only 2.3% compared to clinical OSEM(27). Such speed enhancements are particularly valuable in dynamic LAFOV PET imaging with numerous frames.

An alternative direct AI reconstruction approach for LAFOV PET involves using single slice re-binned (SSRB) 2D sinograms as input for a convolutional encoder-decoder network trained against vendor reconstructions (28). Although reconstruction speed is significantly improved, image quality rapidly deteriorates when inputs fall outside the training distribution. Additionally, the use of SSRB diminishes data fidelity by neglecting oblique lines of responses (LORs) and discarding time-of-flight (TOF) information. Utilizing histo-images or histo-projections instead could better preserve directional information while remaining more compact than full 3D sinograms (29). Deep progressive learning methods have been proposed to integrate the benefits of unrolled iterative and direct AI reconstruction techniques for LAFOV PET (30). This two-stage (denoising followed by enhancement) approach employs unrolled expectation-maximisation iterative reconstructions on uEXPLORER data, surpassing OSEM with 4mm Gaussian filtering. However, it relies on non-convergent target images and requires tracer-specific designs.

Advanced reconstruction methodologies involving diffusion models demonstrated for SAFOV PET (31) have yet to be applied to LAFOV PET datasets. Table 1 summarizes existing AI-enabled PET reconstruction methods for standard and LAFOV PET scanners, highlighting considerations for extending SAFOV methods to LAFOV contexts, particularly addressing challenges related to significantly larger data sizes (32, 33).

*Multiplexed PET (mPET) imaging*

In standard clinical PET workflows, [$^{18}$F]FDG is widely used for metabolic imaging, but [$^{11}$C]methionine, [$^{68}$Ga]/[$^{18}$F] prostate-specific membrane antigen (PSMA) and [$^{68}$Ga]Ga-DOTATATE all offer valuable complementary and clinically relevant information (34). Sequential dual-tracer imaging can provide valuable insights but has limitations, including repeated CT exposure and the need for image registration (34, 35). Multiplexed PET (mPET), where multiple tracers are imaged in a single session via simultaneous or delayed injection and later each tracer's spatial distribution is separated, addresses these challenges.

mPET tracer distributions separation approaches include staggered injections exploiting isotope decay differences (36, 37), and model-based image separation approaches employing tracer kinetic models at the voxel or region-of-interest level (38, 39). AI enhances mPET feasibility and accuracy by learning the individual tracer distributions, to improve their separation accuracy. Applicable methods include direct, model-informed, and unsupervised learning strategies.

Purely data-driven methods use NNs trained either on reconstructed dynamic images (post-separation) (40-42), or directly on mPET sinograms (43, 44) to separate signals. Although effective, these models often require large datasets and may generalize poorly. Incorporating kinetic models improves performance with less data and stronger inductive priors (45). Recent advances combine multi-tracer compartmental models (MTCM) with deep image priors (DIP) for dual-tracer separation without pre-training, reducing bias and variability—though DIP methods may overfit and require high computational resources (46).

Overall, AI is now central to quantitative PET reconstruction, including mPET and LAFOV applications. Even in the absence of high-quality reference data, patient-specific AI models for single- and multi-tracer PET demonstrate improvements over conventional methods, avoiding biases from training distributions.

## AI Methods for Image Resolution Enhancement

Enhanced image resolution in PET imaging is desirable to improve small lesion detection and clinical performance. Resolution in PET is limited by a series of physical factors such as the size of the crystals in (pixelated) detectors, the depth of interaction (DOI), the gamma photon acolinearity and the positron range (PR) of the radionuclides (47). Reconstruction algorithms and management of noise in the image formation process also affect the resolution of the systems as described in the sections dedicated to image reconstruction and denoising in PET. In this section we will focus on the main DL approaches to tackle physical effects affecting resolution with special attention to PR. Last generation clinical PET scans offer a maximum spatial resolution of 2-5 mm (48-50). However, in preclinical PET for small animals such as mice, scanners outperform that resolution and last generation systems can reach up to 1 mm (51). The average positron range in water for 18F, the most used radioisotope in PET is ~0.6 mm (52) posing a lower limit for the resolution of preclinical systems while being far from the typical resolutions attained in clinical scanners. The situation is worsened in the case of other radionuclides like 68Ga (~3.5 mm), 82Rb (~7.1 mm) or 124I (4.4 mm or 2.8 mm for the main two decay modes) affecting considerably the resolution even for clinical scanners (52). PR depends both on the energy of the emitted positrons and on the material travelled by the positron before its annihilation (53, 54). Thus, ranges increase considerably in low density materials such as lung and are reduced in high density materials like bone (55, 56). DL methods can be a promising and interesting application for PR correction, unlike other methodologies either requiring complex computations to properly model the PSF for different combinations of materials and radionuclides (53, 57) or relying on oversimplifications such as constant spread functions across the scanner field-of-view (58, 59), DL can provide an efficient and realistic alternative able to account for the heterogeneous and anisotropic nature of PR. In (60), the authors proposed Deep-PRC, a CNN strategy for PRC for 68Ga in preclinical scanners, recovering resolution compatible with the PSF for 18F. The methodology was later updated to include clinical scanners and to improve the performance of the network (61). CNNs were trained in Monte Carlo (MC) simulations to obtain 68Ga equivalent images from 18F scans. The use of MC simulations is a common strategy for PRC and super resolution in general due to the general lack of real PET datasets of paired (High Resolution) HR and (Low Resolution) LR images. Other works have used a similar strategy to achieve PRC using 68Ga and no range affected images in the simulations (62) or for cardiac PET for $^{82}$Rb (63). In recent works (64), the modelling of the PSF for different radionuclides and materials via DL and its introduction in system response matrix for iterative reconstruction has been proposed showing promising applicability.

Regarding other physical effects affecting resolution in PET, DL has been exploited to enhance PET systems, in particular to reduce parallax errors in DOI identification. For example, in (65) and in (66) the authors proposed the use of CNNs to improve the identification and interaction location of the events within the crystals.

Another important source of resolution loss in PET in the clinical practice comes from patient motion. While most methodologies deal with the motion by compensation and try to correct via gating or frame alignment, for which DL has shown to be efficient and a good candidate to overcome standard approaches (67-69), some other works, such as (70), pose the motion problem as a blurring kernel in the image and solve it as a super-resolution strategy by modeling the kernel motion blur and then training a CNN for resolution recovery, as opposed to applying iterative deconvolution to that motion kernel to effectively deblur the motion-contaminated images as non-ML methods had previously proposed (71).

While the previous methodologies are designed to tackle a specific source of limitations for resolution, there are other generic strategies to generate HR images from LR inputs (72). Some alternatives dedicated to improving image resolution in PET exploit the information of multimodal imaging such as MRI or CT to enhance PET resolution. For example, in (73) the authors proposed a method for super resolution in which CNN were trained with low resolution PET and high-resolution MRI images as input to achieve high-resolution PET images as output. More recently, Gan et al. (74) proposed the use of diffusion methods to generate pseudo-MRI images from PET images which are later used to guide PET reconstruction to deal with partial volume effects and achieve better resolution. Other methods, leverage resolution and noise properties of the system to achieve HR, as a recent study has proposed (75) the use of Fourier Diffusion Models (FDM) to achieve HR PET images from LR inputs.

## Summary

AI-driven methodologies are progressively shaping the future of PET imaging, particularly with the integration of advanced LAFOV PET/CT technologies. These developments promise significant improvements in image reconstruction efficiency, resolution enhancement, multi-tracer imaging, and dynamic imaging. Continued advancements and clinical validations of AI techniques are critical to fully realize their potential, ensuring robust and scalable solutions that address both the technical and practical challenges in modern clinical PET imaging.

## Disclosures

Gary Cook has received research support from Siemens Healthineers. Arman Rahmim is a co-founder of Ascinta Technologies. This research was funded by the Canadian Institutes of Health Research (CIHR) Project Grant (PJT-180251), Canadian Institutes of Health Research (CIHR) Project Grant (PJT-173231). All other co-authors have no relevant conflicts of interest.

39. Sommariva S, Caviglia G, Sambuceti G, Piana M. Mathematical Models for FDG Kinetics in Cancer: A Review. Metabolites. 2021;11(8).
40. Tong J, Wang C, Liu H. Temporal information-guided dynamic dual-tracer PET signal separation network. Medical physics. 2022;49(7):4585-98.
41. Pan B, Marsden PK, Reader AJ. Dual-Tracer PET Image Separation by Deep Learning: A Simulation Study. Applied Sciences. 2023;13(7):4089.
42. Pan B, Marsden PK, Reader AJ. Deep learned triple-tracer multiplexed PET myocardial image separation. Front Nucl Med. 2024;4:1379647.
43. Zeng F, Fang J, Muhashi A, Liu H. Direct reconstruction for simultaneous dual-tracer PET imaging based on multi-task learning. EJNMMI Res. 2023;13(1):7.
44. Wang C, Fang J, Liu H. Direct reconstruction and separation for triple-tracer PET imaging based on three-dimensional encoder-decoder network: SPIE; 2023.
45. Pan B, Marsden PK, Reader AJ. Kinetic model-informed deep learning for multiplexed PET image separation. EJNMMI Phys. 2024;11(1):56.
46. Pan B, Marsden PK, Reader AJ. Self-supervised parametric map estimation for multiplexed PET with a deep image prior. Phys Med Biol. 2025;70(4).
47. Moses WW. Fundamental limits of spatial resolution in PET. Nuclear Instruments and Methods in Physics Research Section A: Accelerators, Spectrometers, Detectors and Associated Equipment. 2011;648:S236-S40.
48. Prenosil GA, Sari H, Fürstner M, Afshar-Oromieh A, Shi K, Rominger A, et al. Performance characteristics of the Biograph Vision Quadra PET/CT system with a long axial field of view using the NEMA NU 2-2018 standard. Journal of nuclear medicine. 2022;63(3):476-84.
49. Spencer BA, Berg E, Schmall JP, Omidvari N, Leung EK, Abdelhafez YG, et al. Performance evaluation of the uEXPLORER total-body PET/CT scanner based on NEMA NU 2-2018 with additional tests to characterize PET scanners with a long axial field of view. Journal of Nuclear Medicine. 2021;62(6):861-70.
50. Yamagishi S, Miwa K, Kamitaki S, Anraku K, Sato S, Yamao T, et al. Performance characteristics of a new-generation digital bismuth germanium oxide PET/CT system, Omni Legend 32, according to NEMA NU 2-2018 standards. Journal of Nuclear Medicine. 2023;64(12):1990-7.
51. McDougald W, Vanhove C, Lehnert A, Lewellen B, Wright J, Mingarelli M, et al. Standardization of preclinical PET/CT imaging to improve quantitative accuracy, precision, and reproducibility: a multicenter study. Journal of Nuclear Medicine. 2020;61(3):461-8.
52. Conti M, Eriksson L. Physics of pure and non-pure positron emitters for PET: a review and a discussion. EJNMMI physics. 2016;3:1-17.
53. Cal-Gonzalez J, Perez-Liva M, Herraiz JL, Vaquero JJ, Desco M, Udias JM. Tissue-dependent and spatially-variant positron range correction in 3D PET. IEEE transactions on medical imaging. 2015;34(11):2394-403.
54. Gavriilidis P, Koole M, Annunziata S, Mottaghy FM, Wierts R. Positron range corrections and denoising techniques for gallium-68 PET imaging: a literature review. Diagnostics. 2022;12(10):2335.
55. Caribé PR, Vandenberghe S, Diogo A, Pérez-Benito D, Efthimiou N, Thyssen C, et al. Monte Carlo simulations of the GE Signa PET/MR for different radioisotopes. Frontiers in Physiology. 2020;11:525575.
56. Kemerink GJ, Visser MG, Franssen R, Beijer E, Zamburlini M, Halders SG, et al. Effect of the positron range of 18 F, 68 Ga and 124 I on PET/CT in lung-equivalent materials. European journal of nuclear medicine and molecular imaging. 2011;38:940-8.
57. Kertész H, Conti M, Panin V, Cabello J, Bharkhada D, Beyer T, et al. Positron range in combination with point-spread-function correction: an evaluation of different implementations for [124I]-PET imaging. EJNMMI physics. 2022;9(1):56.
58. Rukiah A, Meikle SR, Gillam JE, Kench PL, editors. An investigation of 68 Ga positron range correction through de-blurring: A simulation study. 2018 IEEE Nuclear Science Symposium and Medical Imaging Conference Proceedings (NSS/MIC); 2018: IEEE.
10

# TABLES AND FIGURES

*Table 1: Comparison of AI-enabled PET reconstruction methods. Some information adapted from Reader et al. (21).*

| METHOD (YEAR) AND (REF) | GENERAL APPROACH | ARCHITECTURE [NUM. PARAMS] | TRAINING PAIRS USED | SCALABILITY TO LAFOV PET DATA/IMAGES |
|---|---|---|---|---|
| AUTOMAP (2018) (32) | Direct AI [2D] | CNN [~800M] | 50,000 | Possible with SSRB/FORE to create 2D sinograms for training. |
| DEEPPET (2019) (33) | Direct AI [2D] | CED [>60M] | 203,305 | |
| DPIR-NET (2020) (76) | Direct AI [2D] | CED [>60M] | 37,872 | |
| CED LAFOV (2022) (28) | Direct AI [2D] | CED [~64M] | 35,940 | Demonstrated with SSRB sinograms. |
| FASTPET (2021) (26) | Direct AI / Self-supervised [3D] | U-Net [~20M] | 20,297 | Demonstrated with images / histoimages. |
| DPL RECON (2021) (30) | Unrolled [2.5D] | 'FB-NET' [~0.6M] | 161,040 slices (80 subjects) | |
| EM-NET (2019) (77) | Unrolled [3D] | U-Net Shared [~2M] | 16 | Possible with system model considerations. Backpropagating through system model for 3D TB-PET would be difficult on standard workstation hardware. Methodologies where the objective can be split to smaller problems (i.e. view sub-setting) could be a possible solution. Larger networks may be required. |
| MAPEM-NET (2019) (78) | Unrolled [3D] | U-Net not shared [~16M] | 18 | |
| FBSEM-NET (2020) (79) | Unrolled [3D] | U-Net Shared [~77k] | 45 | |
| ITERATIVE NN (2020) (80) | Unrolled [3D] | U-Net | 4 | |

| | | not shared [~40k] | | |
|---|---|---|---|---|
| TRANSEM (2022) (81) | Unrolled [2D] | Swin Transformer [Unspecified] | 510 | |
| DEEP KERNEL (2022) (82) | Deep learned kernel [3D] | Residual U-Net [~2M] | Composite parametric image prior or noisy recons | Tractable with dynamic acquisitions with considerations to the system model. Larger network may be required. |
| NEURAL KEM (2023) (83) | Deep learned kernel [3D] | Modified residual U-Net [~2M] | Composite parametric image prior | |
| DIPRECON (2019) (84) | Self-supervised DIP w/ MRI [3D] | Modified U-Net [~2M] | No prior training needed | Tractable as prior training is not required. Reconstruction speed becomes the bottleneck depending on the deep learning framework used. Deeper models (with higher trainable parameters) may be required. |
| DIP DIRECT PARAMETRIC PET (2022) (85) | Self-supervised DIP w/ MRI [3D] | Modified U-Net [~2M] | No prior training needed | |
| BAYESIAN DEEPRED (2023) (86) | Unsupervised DIP + RED [2D] | U-Nets [~19M & ~1M] | No prior training needed | |
| PET-DDS (2024) (87) | Model informed Diffusion [3D] | Attention U-Net [~40M] | 19 | Network trainable using TB-PET images. Data consistency would require computational time. |
| PET-LISCH (2025) (88) | Model informed Diffusion [3D] | Attention U-Net [~40M] | 55 | |

14